\documentstyle[referee]{laa}

\begin{document}
\thesaurus{07  
	   (07.09.1; 
	    07.13.1;  
	   )}

\title{Electromagnetic Radiation and Motion of Dust Particle -- A Simple Model}
\author{J.~Kla\v{c}ka}
\institute{Institute of Astronomy,
   Faculty for Mathematics and Physics, Comenius University \\
   Mlynsk\'{a} dolina, 842~48 Bratislava, Slovak Republic}
\date{}
\maketitle

\begin{abstract}
A simple model for motion of dust particle (meteoroid) under the action
of (solar) electromagnetic radiation is presented. The particle of the form of
plane mirror is taken into account and exact analytical results are
presented.
As for long-term orbital evolution, particle may spiral outwards the
central body (Sun); initial conditions are important.
As a consequence, motion of real dust particles
may differ from that generally considered.


\end{abstract}

\section{Introduction}
The Poynting-Robertson effect (P-R effect) (Robertson 1937; Kla\v{c}ka 1992a --
the most complete form of the P-R effect) is generally considered to be the
real effect which causes inspiralling of interplanetary dust particles (IDPs),
meteoroids, towards the Sun. (Other, more simple correct derivations may be found
in Kla\v{c}ka's papers: 1992b, 1993a, 1993b). The most general case of the validity
of the P-R effect requires that Eq. (120) (or, Eq. (122) for the moving
particle) in (Kla\v{c}ka 1992a) holds. This may not be the case for the real
nonspherical particle, as it was discussed in Kla\v{c}ka (1993c, 1993d) and
applied in Kla\v{c}ka (1994a).

General equation of motion of IDP in terms of optical properties was presented
by Kla\v{c}ka and Kocifaj (1994) -- the paper does not present any quantitative
calculation.

The aim of this paper is to make a detailed correct calculations for
plane mirror particle as for its equation of motion and as for secular
changes of orbital elements. It is considered that particle rotates around
an axis fixed in space. The advantage of this simple model is that it can
be treated in an analytical way and it may show possible orbital evolution
of real dust particles.

\section{Equation of motion}
Let us consider a particle of a plane form, both sides of which are covered
with mirror. The area of each side is $A$. When the particle's position is
characterized by radius vector $\vec{r}$ with respect to the source of
electromagnetic radiation, it's instantaneous velocity is $\vec{v}$ with
respect to the source.

Density of the flux of the radiation energy of the source is $S$ when
measured in the rest frame of the source. If $c$ is the speed of light
(all occurs in vacuum) and $\hat{\vec{S}} \equiv \vec{r} / |\vec{r}|$,
then the quantities measured in the rest frame of the particle
(proper inertial frame of reference) are:
\begin{eqnarray}\label{1}
\hat{\vec{S'_{i}}} &=& ( 1 ~+~ \vec{v} \cdot \hat{\vec{S}} / c ) ~
		       \hat{\vec{S}}   ~-~ \vec{v} / c ~,
\nonumber \\
\hat{\vec{S'_{o}}} &=& \hat{\vec{S'_{i}}} ~-~ 2 ~ (
		       \hat{\vec{n'}} \cdot \hat{\vec{S'_{i}}} ) ~
		       \hat{\vec{n'}}  ~,
\nonumber \\
S' &=& ( 1 ~-~ 2~ \vec{v} \cdot \hat{\vec{S}} / c ) ~ S ~,
\end{eqnarray}
where unit vector $\hat{\vec{S'_{o}}}$ represents direction and orientation
of the beam reflected from the plane mirror characterized by normal unit
vector $\hat{\vec{n'}}$; all quantities are correct to the first order
in $v/c$.

If the plane mirror is turned at an angle $\delta$ about an axis
characterized by unit vector $\vec{a}$, the new normal unit vector
$\hat{\vec{n'_{n}}}$ (coordinate system is fixed -- it does not change) is
\begin{equation}\label{2}
\hat{\vec{n'_{n}}} =  \hat{\vec{n'}} ~ \cos \delta ~+~
		      \vec{a} ~ ( \vec{a} \cdot  \hat{\vec{n'}} ) ~
		      ( 1 ~-~ \cos \delta ) ~+~
		      ( \vec{a} \times	\hat{\vec{n'}} ) ~ \sin \delta ~.
\end{equation}

We will consider that
$\hat{\vec{n'}} = \pm \hat{\vec{x'}} = \pm \hat{\vec{x}}$ and the particle
rotates around axis $\hat{\vec{x'}}$, for the sake of simplicity.
The effective area for the incident
radiation is
\begin{equation}\label{3}
A_{eff} = A ~ | \cos \Theta | ~,
\end{equation}
where $\Theta$ is the angle between $\hat{\vec{S}}$ and $\hat{\vec{x}}$
($\vec{r} = x~ \hat{\vec{x}} ~+~ y ~ \hat{\vec{y}}~+~ z ~ \hat{\vec{z}}$).
We can write, thus:
\begin{eqnarray}\label{4}
\hat{\vec{n'}} &=& -~ \hat{\vec{x'}} = -~ \hat{\vec{x}} ~, ~~
\Theta \in ~ < -~ \pi / 2, \pi / 2 >	~,
\nonumber \\
\hat{\vec{n'}} &=& +~ \hat{\vec{x'}} = +~ \hat{\vec{x}} ~, ~~
\Theta \in ~ < \pi / 2, 3~ \pi / 2 >	~.
\end{eqnarray}

Equation of motion for the particle has the following form
\begin{eqnarray}\label{5}
\frac{d \vec{p'}}{dt} &=&
\frac{A_{eff} ~ S'}{c} ~ \left ( \hat{\vec{S'_{i}}} ~-~
				 \hat{\vec{S'_{o}}} \right ) ~,
\nonumber \\
\frac{d E'}{dt} &=& 0
\end{eqnarray}
in the proper inertial frame of reference.

Using the previous relations (and Lorentz transformation to the first
order in $v/c$), we finally obtain equation of motion in the reference
frame of the source of electromagnetic radiation:
\begin{equation}\label{6}
\dot{\vec{v}} = 2 ~ \frac{A~S}{m~c} ~ \left \{
\left ( 1 ~-~ \frac{\vec{v}}{c} \cdot \hat{\vec{S}} \right ) ~
	      \hat{\vec{S}} \cdot \hat{\vec{x}} ~-~
\frac{\vec{v}}{c} \cdot \hat{\vec{x}} \right \} ~ | \cos \Theta | ~
	      \hat{\vec{x}} ~,
\end{equation}
where $m$ is mass of the particle and the dot denotes differentiation
with respect to time.

\section{Electromagnetic radiation and gravitation}
Taking into account gravitational acceleration of the central body of mass $M$,
we can write the complete equation of motion in the form
\begin{eqnarray}\label{7}
\dot{\vec{v}} &=& -~ \frac{\mu}{r^{3}} ~ \vec{r} ~+~
		2 ~ \beta ~ \frac{\mu}{r^{2}} ~ \left \{
\left ( 1 ~-~ \frac{\vec{v}}{c} \cdot \hat{\vec{S}} \right ) ~
	      \hat{\vec{S}} \cdot \hat{\vec{x}} ~-~
\frac{\vec{v}}{c} \cdot \hat{\vec{x}} \right \} ~ | \cos \Theta | ~
	      \hat{\vec{x}} ~,
\nonumber \\
\beta ~ \frac{\mu}{r^{2}} &=& \frac{A~S}{m~c} ~,
\end{eqnarray}
where $\mu = G~M$ and $G$ is gravitational constant.

\section{Secular changes of orbital elements}
Let the initial position and velocity vectors lie in the $xy-$plane.
Eq. (7) yields that the orbital plane conserves. Secular changes of other
orbital elements can be also calculated in an analytical way. The
results are summarized in Eqs. (8) -- (10) for semimajor axis $a$,
eccentricity $e$ and longitude of pericenter $\omega$
(the quantity $p$ is defined $p = a~ ( 1 ~-~ e^{2} )$).
\begin{equation}\label{8}
< \frac{d ~a}{d ~t} > ~=~ \frac{4}{\pi} ~ \frac{\beta ~\mu}{c} ~
  \frac{2~/~3 ~+~ \left [ 6 ~ \cos \left ( 2 ~ \omega \right ) ~ / ~ 5
   ~-~ 1 \right ]~ e^{2}}{a~ \left ( 1 ~-~ e^{2} \right )^{3/2}}
\end{equation}
\begin{eqnarray}\label{9}
< \frac{d ~e}{d ~t} > ~ &=& ~ \frac{2}{\pi} ~ \frac{\beta ~\sqrt{\mu}}{a^{3/2}} ~
  \left \{ \frac{4}{15} ~e ~ \left ( 1 ~-~ e^{2} \right ) ~
  \sin \left ( 2~ \omega \right ) ~+ \right .
\nonumber \\
& & \left . +~
\left ( 1 ~-~ e^{2} \right ) ~
  \sum_{n=2}^{\infty} ~ e^{2n-1} ~ \int_{- ~\pi / 2}^{\pi / 2} ~
  \cos ^{2} \Theta  ~ \sin \Theta \cos^{2n} \left ( \Theta ~-~ \omega \right ) ~
  d ~ \Theta ~+~ \right .
\nonumber \\
& & \left . + ~ \frac{\sqrt{\mu ~/~ p}}{c} ~e ~ \left [
    \frac{2}{5} ~-~ \frac{22}{15} \sin ^{2} \omega ~+~
    \frac{4}{15} ~ \cos  \left ( 2 ~ \omega \right ) ~+~ \right . \right .
\nonumber \\
& & \left .  \left . +~ 2 ~
    \left ( 1 ~-~ e^{2} \right ) ~ \int_{- ~\pi / 2}^{\pi / 2} ~
    \frac{\cos ^{2} \Theta ~ \sin \Theta ~ \cos \left ( \Theta ~-~ \omega \right )
    ~ \sin \left ( \Theta ~-~ \omega \right )}{1 ~-~
    e^{2} ~\cos^{2} \left ( \Theta ~-~ \omega \right )} ~ d ~ \Theta
    \right ] \right \}
\end{eqnarray}
\begin{eqnarray}\label{10}
< \frac{d ~\omega}{d ~t} > ~ &=& ~ -~ \frac{2}{\pi} ~ \frac{\beta ~\sqrt{\mu}}{a^{3/2} ~e} ~
  \left \{ \frac{2}{3} ~
  \cos ~ \omega  ~+ \right .
\nonumber \\
& & \left . +~
    \sin \left ( 2~ \omega \right ) ~ \sin ~ \omega ~
    \sum_{n=0}^{\infty} ~ \left ( e ~ \sin \omega \right ) ^{2n}
    \left [ \frac{1}{2~n ~+~ 1} ~+~
    \frac{\cos \left ( 2 ~ \omega \right )}{2~n ~+~ 3} \right ] ~+ \right .
\nonumber \\
& & \left . +~ \cos \left ( 2~ \omega \right ) ~
    \sum_{n=0}^{\infty} ~ \frac{e^{2n}}{2~n ~+~ 3} ~
~ \int_{- ~\pi / 2 ~-~ \omega}^{\pi / 2 ~-~ \omega} ~
  \cos ^{2n+1} x ~ d~ x ~- \right .
\nonumber \\
& & \left . - ~ \frac{\sqrt{\mu ~/~ p}}{c} ~e ~ \left [
	\frac{3}{5} ~ \sin \left ( 2 ~ \omega \right )
    ~+~ 4 ~ \sin ^{3} \omega ~ \cos \left ( 2 ~\omega \right ) ~
    \sum_{n=0}^{\infty} ~ \frac{\left ( e~\sin \omega \right )^{2n}}{2~n ~+~ 3} ~
    \right ] \right \}
\end{eqnarray}
Initial values
for osculating orbital elements must be calculated for the case
when central acceleration is given by the value $\mu$ -- as for comparison
see Kla\v{c}ka (1992c, 1999).

\section{Discussion to secular changes of orbital elements}
Eq. (10) shows that advance of pericenter (perihelion, in the case of Solar
System) exists. If the relativistic terms proportional to
$\sqrt{\mu/p} ~/~ c$ would not exist, then $< d \omega / d t >$ would be
zero for $\cos \omega = 0$, i. e. for $\omega = \pm \pi/2$. The relativistic
terms yield that $< d \omega / d t >$ $< 0$ for $\omega = + \pi/2$ and
$< d \omega / d t >$ $> 0$ for $\omega = - \pi/2$.

The stable situation occurs at $\omega = - \pi/2 ~+~ \varepsilon$, where
$\varepsilon$ is small positive number (depends on the value of eccentricity).
Eq. (9) yields that $< d e / d t >$ $< 0$. Eq. (8) yields that
$< d a / d t >$ $< 0$ for $e > \sqrt{10/33}$ (approximately) and
$< d a / d t >$ $> 0$ for $e < \sqrt{10/33}$ (approximately). Thus, if the
initial eccentricity is greater than $e > \sqrt{10/33}$ (approximately), the
particle spiral towards the central body, its eccentricity still decreases.
When $e < \sqrt{10/33}$ (approximately) occurs, the particle
spiral outwards the central body.

%

\section{Conclusion}
We have shown, in an analytical way, that special type of particle
exhibits orbital evolution which is not consistent with that for the
Poynting-Robertson effect (moreover, the behaviour depends on initial
conditions).
Therefore, the P-R effect cannot be simply applied to the study
of motion of real particles.
(Numerical calculations for real particle are in preparation.)

\acknowledgements
Special thanks to the firm ``Pr\'{\i}strojov\'{a} technika, spol. s r. o.''.
The paper was partially
supported by the Scientific Grant Agency VEGA (grant No. 1/7067/20).

\end{document}